\documentclass[reprint,amsmath,amssymb,aps,prl,nofootinbib]{revtex4-2}
\usepackage{endnotes}
\usepackage{graphicx}
\usepackage{dcolumn}
\usepackage{bm}
\usepackage{hyperref}
\usepackage{graphicx,amssymb,amsmath,amsfonts}
\usepackage{verbatim}

\def\lvec#1{\setbox0=\hbox{$#1$}
    \setbox1=\hbox{$\scriptstyle\leftarrow$}
    #1\kern-\wd0\smash{
    \raise\ht0\hbox{$\raise1pt\hbox{$\scriptstyle\leftarrow$}$}}
    \kern-\wd1\kern\wd0}
\def\rvec#1{\setbox0=\hbox{$#1$}
    \setbox1=\hbox{$\scriptstyle\rightarrow$}
    #1\kern-\wd0\smash{
    \raise\ht0\hbox{$\raise1pt\hbox{$\scriptstyle\rightarrow$}$}}
    \kern-\wd1\kern\wd0}

\def\diracstar#1#2{
    \setbox0=\hbox{$\gamma$}\setbox1=\hbox{$\gamma_{#1}$}\includegraphics[]
    \gamma_{#1}\kern-\wd1\kern\wd0
    \smash{\raise4.5pt\hbox{$\scriptstyle#2$}}}

\newcommand{\beq}{\begin{equation}}
\newcommand{\eeq}{\end{equation}}
\newcommand{\beqn}{\begin{eqnarray}}
\newcommand{\eeqn}{\end{eqnarray}}
\newcommand{\nn}{\nonumber}

\begin{document}
\makebox[0pt][r]{\text{ITCP-IPP 2025/1}\hspace{-8cm}}\vspace{0.1 cm}

\makebox[0pt][r]{\text{CCTP-2025-1}\hspace{-8cm}}\vspace{.1 cm}

\title{ A model calculation of the CKM matrix}

\author{Maurizio Firrotta}
\email{mfirrotta@physics.uoc.gr}
 
\affiliation{Crete Center for Theoretical Physics, Institute for Theoretical and Computational Physics\\
Department of Physics, Voutes University Campus, GR-70013, Heraklion, Greece
}
\author{Giancarlo Rossi}
\email{rossig@roma2.infn.it}
\affiliation{
Universit\`a di Roma
  {\it Tor Vergata} and INFN Sezione di Roma 2, Via della Ricerca Scientifica - 00133 Roma, Italy\\ Centro Fermi - Museo Storico della Fisica e Centro Studi e Ricerche ``E.\ Fermi''\\
Piazza del Viminale 1 - 00184 Roma, Italy}

\date{\today}
\begin{abstract}
We propose a strategy to compute the CKM matrix based on the conjecture, recently put forward in the literature, according to which elementary particle masses are not generated like in the standard Higgs scenario, but emerge from a non-perturbative mechanism triggered by the presence in the fundamental Lagrangian of ``irrelevant'' chiral breaking operators of the Wilson type of dimension $d\geq 6$ scaled by $d-4$ powers of the UV cutoff. Non-perturbatively generated quark masses have the form $m_q\sim C_q(\alpha) \Lambda_{RGI}$ where $\Lambda_{RGI}$ is the RGI scale of the theory and $C_q(\alpha)$ is a function of the gauge couplings. For the (elementary) fermion $q$ the $C_q(\alpha)$ leading behaviour is $C_q(\alpha)={\mbox{O}}(\alpha^{1+(d_q-4)/2})$. The dependence of the gauge coupling power behaviour from the dimension $d_q$ of the Wilson-like operators associated with the fermion $q$ can be exploited to construct hierarchically organized up and down ``proto-mass matrices'' for ``proto-flavours'', the diagonalization of which yields flavoured quarks with definite masses and a first principle construction of the CKM matrix.
\end{abstract}

\maketitle

\section{Introduction}
\label{sec:INTRO}
Based on the results of~\cite{Frezzotti:2014wja}, where it is shown that in certain strongly interacting gauge theories the mass of the fundamental fermion can be non-perturbatively (NP-ly) generated without any Higgs mechanism in place, in ref.~\cite{Rossi:2024fku} (see also~\cite{Rossi:2023xhl,Rossi:2023wok}) the construction of realistic beyond-the-Standard-Model models (bSMm's) the low energy limit of which is the SM, was developed. 

In this NP scenario elementary particle masses will be proportional to the RGI scale of the theory, $\Lambda_{RGI}$ (and not to the vev of the Higgs field) times coefficient functions depending on the gauge couplings.

The distinctive feature of these unconventional models is that chiral symmetry is broken at the UV scale by ``irrelevant'' gauge invariant $d\geq 6$ operators of the Wilson type~\footnote{By this we mean operators of dimension larger than 4, scaled by powers of the UV cutoff (like in the case of the Wilson term in lattice QCD) and not by some external mass scale.}. In the Wigner phase of the theory, however, chiral symmetry can be recovered at low energy by the ``natural''~\cite{THOOFT} tuning of certain Lagrangian parameters. In the Nambu--Goldstone phase at this critical point, dynamical mass terms for elementary particles can be shown to be NP-ly generated~\cite{Frezzotti:2014wja,Rossi:2024fku}.\ They appear as a kind of NP anomalies preventing full recovery of chiral symmetry and emerge from a delicate interplay between UV loop behaviour and IR features originating from the spontaneous breaking of the (recovered) chiral symmetry. Models of this kind have a number of interesting features which we list here for completeness~\cite{Frezzotti:2014wja,Rossi:2024fku,Frezzotti:2016bes}.

1) There is no problem with the tuning of the Higgs mass as there is no Higgs boson.

2) At the critical point where chirality is recovered, the theory does not suffer from a strong CP problem.

3) To get the correct order of magnitude of the {\it top} quark mass we are led to conjecture the existence of a new sector of strongly interacting particles (Tera-particles~\cite{Glashow:2005jy}) gauge invariantly coupled to standard matter, living at the TeV scale. The electroweak (EW) scale is interpreted as the RGI scale of the new Tera interaction.

4) The masses of the elementary particles are no longer free parameters, but are determined by the NP dynamics of the theory.

5) The low energy Lagrangian of the model, obtained after integrating out the particles of the heavy Tera sector, looks like the SM. 

6) The full theory, including SM degrees of freedom and the particles of the new sector, displays unification of gauge couplings.

Taking inspiration from this unconventional theoretical framework, in the present note we propose a scheme that allows a physically motivated understanding of the structure of the CKM matrix~\cite{Cabibbo:1963yz,Kobayashi:1973fv} and gives some clue about the origin of the notion of flavor. 

From a different vantage point, interesting models giving hints on the nature of flavour and the pattern of the CKM matrix have been recently proposed and constructed in refs.~\cite{Barbieri:2023qpf,Barbieri:2024zkh}.

Technically the present approach is based on exploiting the liberty that one has in the kind of bSMm's developed in refs.~\cite{Frezzotti:2014wja,Rossi:2024fku} to adjust the dimension of the quark Wilson-like operators so as to match the order of magnitude of the various quark mass terms. It can be proved, in fact, that the larger is the dimension of the Wilson-like operator associated with a fermion, the higher will be the power of the gauge coupling in front of its NP self-energy diagrams, hence generically the smaller the magnitude of the correspondingly generated mass term. In this way, Wilson-like operators with larger and larger dimension (6 is the minimal allowed value of the dimension of such operators) generate smaller and smaller mass terms. Naturally, in the presence of an entangled network of Wilson-like operators mixing will be a key issue to identify the physical quark states with definite mass and hence flavour. Solving this mixing will be the main topic of the next section.

\section{A strategy for the calculation of the CKM matrix}
\label{sec:SCCKM}

The main idea behind the construction of the CKM matrix we are going to present is the assumption that at the fundamental level there exist ``proto-quarks'' for which Wilson-like terms are hierarchically ``proto-flavour blind''. To precisely see what we mean by this, let us start discussing the case of the up quarks. The discussion for the down quarks will be totally analogous. In this note we neglect $CP$ violation effects.

We will assume that the mass matrix of the up quarks generated by the Wilson-like operators in the fundamental Lagrangian through the NP mechanism described in~\cite{Frezzotti:2014wja,Rossi:2024fku} has the expression~\footnote{Formally this construction can be easily extended to any number of flavours, although eventually   asymptotic freedom will be lost.} 
\beqn
M^{u}=C_{q^u}^{(3)}{\cal M}_{3I}+  C_{q^u}^{(2)}{\cal M}_{2I}+ C_{q^u}^{(1)}{\cal M}_{1I}
\label{MUP}
\eeqn
where ${\cal M}_{3I}$, ${\cal M}_{2I}$ and ${\cal M}_{1I}$ are matrices of the form 
\begin{widetext}
\small
\begin{equation}
{\cal M}_{3I}=\left(\begin{array}{ccc}
1 & 1& 1 \\
1 & 1& 1 \\
1 & 1 & 1
\end{array}
\right)\,,\quad   {\cal M}_{2I}=\left(\begin{array}{ccc}
0 & 0 &  0 \\
0 &1 &1\\
0 & 1 &1
\end{array}
\right)\,, \quad {\cal M}_{1I}=\left(\begin{array}{ccc}
0 & 0 &  0 \\
0 &0 &0\\
0 & 0 &1
\end{array}\right)\, .
\end{equation}
\end{widetext}
The first term in eq.~(\ref{MUP}) comes from Wilson-like operators of $d=6$. According to the analysis developed in refs.~\cite{Frezzotti:2014wja,Rossi:2024fku}, it will have the gauge coupling scaling behaviour 
{\small\beqn
C_{q^u}^{(3)}(\alpha_s,\alpha_Y;y_{q^u}) =C^{(3)}_u(\alpha)\Lambda_{RGI} \, , \,\, \,C^{(3)}_u(\alpha)= {\mbox{O}}(\alpha^2)\label{SCU1}\, ,
\eeqn}
where $\alpha$ can be either $\alpha_s$ or $\alpha_Y$ and $y_{q}$ is the hypercharge of the ``proto-quark'' $q^u$. At this stage, we do not commit ourselves to any special value of the gauge couplings. 

The first matrix in $M^u$ comes from maximally ``proto-flavour'' violating $d=6$ Wilson-like operators. If we would limit to these operators, the diagonalization of $M^u$ would give one ``flavour'' eigenstate with eigenvalue $3\, C_{q^u}^{(3)}$ and two degenerate ones with vanishing eigenvalue. However, it is clear that we can continue by introducing Wilson-like operators of $d=8$ and $d=10$ leading to higher orders proto-flavour contributions so as to give hierarchically ordered, non-vanishing and non-degenerate masses to the next two flavour eigenvectors. In summary the basic up quark mass matrix, before diagonalization, will have the form displayed in~(\ref{MUP}) with $C_{q^u}^{(3)}$ given by eq.~(\ref{SCU1}) and, according to the scaling rule $C(\alpha)={\mbox{O}}(\alpha^{1+(d-4)/2})$~\cite{Rossi:2024fku}
{\small\beqn
\hspace{-.4cm}&&C_{q^u}^{(2)}(\alpha_s,\alpha_Y;y_{q^u})=C^{(2)}_u(\alpha)\Lambda_{RGI} \, , \,\, \,\, C^{(2)}_u(\alpha)= {\mbox{O}}(\alpha^3)\, ,\label{SCU31}\\
\hspace{-.4cm}&&C_{q^u}^{(1)}(\alpha_s,\alpha_Y;y_{q^u})=C^{(1)}_u(\alpha)\Lambda_{RGI} \, , \,\,\,\, C^{(1)}_u(\alpha)= {\mbox{O}}(\alpha^4)\label{SCU41}
\eeqn}
in correspondence to Wilson-like operators of $d=8$ and $d=10$, respectively.

Diagonalization of the $M^u$ matrix identifies the up quark flavour states with definite mass. In the next subsection we will give the explicit expression of the matrix $V^u$ that diagonalizes $M^u$ and of the eigenvalues $\lambda_i^u, i=1,2,3$, which will represent the masses of the up quarks with definite flavour. Naturally, we will call {\it top} the eigenstate corresponding to the largest eigenvalue, {\it charm} the eigenstate corresponding to the second largest and {\it up} the eigenstate corresponding to the smallest eigenvalue.

Similarly to what we did for $M^u$ we write for $M^d$
\begin{eqnarray}
 M^{d}= C_{q^d}^{(3)}{\cal M}_{3I}+ C_{q^d}^{(2)}{\cal M}_{2I}+  C_{q^d}^{(1)}{\cal M}_{1I}\, ,
\label{MDOWN}
\end{eqnarray} 
where we assume for the $C_{q^d}^{(i)}, i=1,2,3$ coefficients the gauge coupling scaling behaviour
{\small
\beq
C_{q^d}^{(3)}(\alpha_s,\alpha_Y;y_{q^d})=C^{(3)}_d(\alpha)\Lambda_{RGI} \, , \,\, C^{(3)}_d(\alpha)= {\mbox{O}}(\alpha^3)\, ,\label{SCD21}
\eeq
\beq
C_{q^d}^{(2)}(\alpha_s,\alpha_Y;y_{q^d})=C^{(2)}_d(\alpha)\Lambda_{RGI} \, , \,\, C^{(2)}_d(\alpha)= {\mbox{O}}(\alpha^4)\, ,\label{SCD31}
\eeq
\beq
C_{q^d}^{(1)}(\alpha_s,\alpha_Y;y_{q^d})=C^{(1)}_d(\alpha)\Lambda_{RGI} \, , \,\, C^{(1)}_d(\alpha)= {\mbox{O}}(\alpha^5)\label{SCD41}\, .
\eeq}
Comparing eqs.~(\ref{SCU1}), (\ref{SCU31}) and~(\ref{SCU41}) with eqs.~(\ref{SCD21}), (\ref{SCD31}) and~(\ref{SCD41}), respectively, we see that the gauge coupling scaling power of the down coefficients, $C_{d}^{(i)}, i=1,2,3$, is larger by one unit with respect to that of the corresponding up coefficients, $C_{u}^{(i)}, i=1,2,3$. Admittedly, this is an {\it ad hoc} assumption that we introduce to deal with the fact that generically down quarks are lighter than up quarks. In summary at the Lagrangian level this means that the Wilson-like terms associated with the up quarks will be $d=6$, $d=8$, and $d=10$ operators, while those associated with down quarks will be $d=8$, $d=10$, and $d=12$ operators.
\vspace{-1cm}
\subsection*{The mass matrices}
\label{sec:DIA}
Introducing for up and down sectors the definitions
\beqn
\rho_{23}^ {u,d}\equiv\frac{C^{(2)}_{q^{u,d}}}{C^{(3)}_{q^{u,d}}},\quad
\rho_{13}^{u,d}\equiv\frac{C^{(1)}_{q^{u,d}}}{C^{(3)}_{q^{u,d}}}\label{R31UD}
 \eeqn
 with $\rho_{23}^ {u,d}={\mbox{O}}(\alpha)$ and $\rho_{13}^ {u,d}={\mbox{O}}(\alpha^2)$,
we can write the matrices $M^u$ and $M^d$ in the compact and convenient form
\begin{eqnarray}
\hspace{-1.cm}&&M^{u,d}=C^{(3)}_{q^{u,d}} \left(
\begin{array}{ccc}
1 & 1& 1 \vspace{1mm}\\
1 & 1+\rho_{23}^{u,d} & 1+\rho_{23}^{u,d}  \vspace{1mm}\\
1 & 1+\rho_{23}^{u,d}  & 1+\rho_{23}^{u,d} +\rho^{u,d}_{13}
\end{array}
\right)  
 \label{MTRIAL2NUD}
 \end{eqnarray}  
\subsection*{The CKM matrix}
\label{sec:CKMMA}
 
We now proceed to the explicit diagonalization of the matrices $M^u$ and $M^d$. We call $[\vec \chi^{\,u}_f, \lambda_f^u, f=1,2,3]$ (eventually identified with $u,c,t$) and $[\vec\chi^{\,d}_f, \lambda_f^d, f=1,2,3]$ (eventually identified with $d,s,b$) the eigenvectors and the eigenvalues of the matrices $M^u$ and $M^d$, respectively. We recall that, since $M^u$ and $M^d$ are hermitian matrices, they have real eigenvalues (we neglect CP violation effects) and orthogonal eigenvectors. From the knowledge of the eigenvectors one can construct the unitary matrices $V^u$ and $V^d$ which allow us to write
\beqn
M^u=V^u \lambda^u V^{u\dagger}\, ,\label{VU}\, \quad M^d=V^d \lambda^d V^{d\dagger}\, ,
\label{VD}
\eeqn
and finally compute the CKM matrix from~\cite{CKMM}
\beq
V_{\rm CKM} = V^{u\dagger}V^d\, .
\label{CKMM}
\eeq
\vspace{-1 cm}
\subsection*{Eigenvectors and eigenvalues}
\label{sec:EAE}

The exact algebraic expressions of the eigenvalues and eigenvectors of $M^d$ and $M^d$ are quite involved, but can be worked out. Making explicit the gauge coupling scaling behaviour suggested by the eqs.~(\ref{R31UD}) and defining 
\begin{equation}
\rho_{23}^{u,d}=\alpha\, r_{23}^{u,d}\,,\quad \rho_{13}^{u,d}=\alpha^{2} r_{13}^{u,d} \, ,\label{DEFR}
\end{equation}
with $r_{23}^{u,d}$ and $r_{13}^{u,d}$ O(1) coefficients, it is instructive to expand the exact results perturbatively in $\alpha$. Making use of the scaling~(\ref{DEFR}), for the $M^{u,d}$ matrices eigenvalues one finds the expansions
{\small\beq
\lambda^{u,d}_{1}=\frac{\alpha^{2}}{2}C_{q^{u,d}}^{(3)} r_{13}^{u,d}+{\mbox{O}}(\alpha^5)\label{EVAU}
\eeq

\beq
\lambda^{u,d}_{2}=C_{q^{u,d}}^{(3)}\Bigg[\frac{2\alpha}{3} r^{u,d}_{23} - \frac{\alpha^2}{6} \Big(\frac{16}{9} (r^{u,d}_{23})^2 {-}  r^{u,d}_{13}\Big)\Bigg]+{\mbox{O}}(\alpha^5)\label{EVAC}
\eeq

\beq
\lambda^{u,d}_{3}\!=\! C_{q^{u,d}}^{(3)} \!\Bigg[3+ \!\frac{4\alpha}{3}  r^{u,d}_{23} \!+ \!\frac{\alpha^2}{3}  \Big(\frac{8}{9} (r^{u,d}_{23})^2 \!+\!  r_{13}^{u,d}\Big)\!\Bigg]+\!{\mbox{O}}(\alpha^5)\label{EVAT}
\eeq}
Observing that the eigenvalue are hierarchically ordered in $\alpha$, we are in a natural way lead to the identifications
\beqn
&&\lambda^{u}_{1}={\mbox{O}}(\alpha^4)\to m_{u}\,,\quad \lambda^{u}_{2}={\mbox{O}}(\alpha^3) \to m_{c}\nonumber\\ &&\hspace{2cm}\lambda^{u}_{3}={\mbox{O}}(\alpha^2)\to m_{t} ,\label{IDMUCT}\\
&&\lambda^{d}_{1}={\mbox{O}}(\alpha^5) \to m_{d}\,,\quad \lambda^{d}_{2}={\mbox{O}}(\alpha^4) \to m_s \nonumber \\ 
&&\hspace{2cm}\lambda^{d}_{3}={\mbox{O}}(\alpha^3) \to m_{b}\, .
\label{IDMDSB}
\eeqn
Recalling the gauge coupling scaling behaviours~(\ref{SCU1})--(\ref{SCU41}) and~(\ref{SCD21})--(\ref{SCD41}), one gets the expansions 
{\small
\beqn
\hspace{-1.8cm}&& m_{u}= {1\over 2} C_{q^{u}}^{(1)}+{\mbox{O}}(\alpha^5) ={1\over 2} C_{q^{u}}^{(3)} \rho^{u}_{13}+{\mbox{O}}(\alpha^5) \label{ID1}\\
\hspace{-1.8cm}&& m_{c}= {2\over 3} C_{q^{u}}^{(2)}+{\mbox{O}}(\alpha^4) ={2\over 3} C_{q^{u}}^{(3)}\rho^u_{23}+{\mbox{O}}(\alpha^4) \label{ID2}\\
\hspace{-1.8cm}&& m_{t}= 3 C_{q^{u}}^{(3)}+{\mbox{O}}(\alpha^3) \label{ID3}\\
\hspace{-1.8cm}&& m_{d}= {1\over 2} C_{q^{d}}^{(1)} +{\mbox{O}}(\alpha^6)={1\over 2} C_{q^{d}}^{(3)}\rho^d_{13}+{\mbox{O}}(\alpha^6)\label{ID4}\\
\hspace{-1.8cm}&& m_{s} ={2\over 3}\, C_{q^{d}}^{(2)}+{\mbox{O}}(\alpha^5)={2\over 3} C_{q^{d}}^{(3)}\rho^d_{23}+{\mbox{O}}(\alpha^5)  \label{ID5}\\
\hspace{-1.8cm}&&m_{b}= 3C_{q^{d}}^{(3)}+{\mbox{O}}(\alpha^4)\label{ID6}
\eeqn}
In ref.~\cite{Rossi:2024fku} it is shown that the mass formulae~(\ref{ID3}) and~(\ref{ID6}) provide rather good estimates of the top and bottom quark mass in units of $M_W$. 

Neglecting O($\alpha^3$) terms, one finds for the normalized eigenvectors 
\vspace{-0.5cm}
\begin{widetext}
\begin{equation}
\vec{\chi}_{1}=
{1\over \sqrt{2}}\begin{pmatrix}
{\alpha\over 2}{r_{13}\over r_{23}}+ {\alpha^{2}\over 4} {r_{13}^{2}\over r_{23}^{2}}\\
 -{1}-{\alpha\over 4}{r_{13}\over r_{23}}-{\alpha^{2}\over 32 }{r_{13}^{2}\over r_{23}^{2}}\\
  {1}-{\alpha\over 4}{r_{13}\over r_{23}}-{7\alpha^{2}\over 32}{r_{13}^{2}\over r_{23}^{2}}
\end{pmatrix}+\dots \quad {\vec{\chi}_{2}=
{1\over \sqrt{6}}\begin{pmatrix}
-2-{4\over 9}\alpha r_{23}+ {\alpha^{2}\over   r_{23}^{2}}(\frac{16}{81} r_{23}^{4}\!-\!\frac{1}{9} r_{23}^{2}r_{13}\!+\!\frac{3}{16} r_{13}^{2})\vspace{1mm}\\ 
\!{1}\!-\!{\alpha\over  r_{23}}({4\over 9} r_{23}^{2}\!+\! {3\over 4} r_{13})\!+\! {\alpha^{2}\over r_{23}^{2}}(\frac{4}{81} r_{23}^{4}\!-\!\frac{1}{9} r_{23}^{2} r_{13}\!-\!\frac{15}{32} r_{13}^{2})\vspace{1mm}\\
\!{1}\!-\!{\alpha\over  r_{23}}({4\over 9} r_{23}^{2}\!-\!{3\over 4} r_{13})\!+\! {\alpha^{2}\over  r_{23}^{2}}(\frac{4}{81} r_{23}^{4}\!-\!\frac{1}{9} r_{23}^{2} r_{13}\!+\!\frac{9}{32} r_{13}^{2})
\end{pmatrix}}+\dots\label{EVCUCC}
\end{equation}
\end{widetext}
\begin{equation}
\vec{\chi}_{3}\!=
{1\over\sqrt{3}}\begin{pmatrix}
{1} - {4\over 9 } \alpha\, r_{23} + 
  \alpha^2 ({4\over 81} r_{23}^2 - {1\over 9} r_{13})\vspace{1mm}\\
 {1} + {2\over 9 } \alpha\, r_{23} -
  \alpha^2 ({8\over 81}  r_{23}^2 + {1\over 9} r_{13})\vspace{1mm}\\
 {1} + {2\over 9} \alpha\, r_{23} -  \alpha^2 ({8\over 81} r_{23}^2 - {2\over9} r_{13})
\end{pmatrix}{+}\dots\label{EVCT}
   \end{equation}
These equations express the physical quark flavour eigenstates in the proto-flavour basis. Up and down eigenstates are obtained from the general formulae~(\ref{EVCUCC})--(\ref{EVCT}) with the replacements $r_{23}\to r_{23}^u, r_{13}\to r_{13}^u$ and $r_{23}\to r_{23}^d, r_{13}\to r_{13}^d$, respectively, leading to the identifications
\beq 
\hspace{-.6cm}\vec{\chi}^{\,u}_{1}=\vec{\chi}_{up}\,,\,\,\,\quad \vec{\chi}^{\,u}_{2}=\vec{\chi}_{charm}\,,\,\,\quad \vec{\chi}^{\,u}_{3}=\vec{\chi}_{top}
\eeq
\beq 
\vec{\chi}^{\,d}_{1}=\vec{\chi}_{down}\,,\qquad \vec{\chi}^{\,d}_{2}=\vec{\chi}_{strange}\,,\quad \vec{\chi}^{\,d}_{3}=\vec{\chi}_{bottom}
\eeq
In terms of these normalized eigenstates one can now construct the diagonalizing unitary matrices, $V^u$ and $V^d$. They are given by the formulae
{\small \begin{equation}
V^{u}=(\vec{\chi}_{up}\downarrow \, \vec{\chi}_{charm}\downarrow\, \vec{\chi}_{top}\downarrow)
\end{equation}
\begin{equation}
V^{d}=(\vec{\chi}_{down}\downarrow \, \vec{\chi}_{strange}\downarrow\, \vec{\chi}_{bottom}\downarrow)
\end{equation}
The CKM matrix is defined by the relation 
\begin{equation}
V_{CKM}=V^{u\dagger}V^{d}=\begin{pmatrix}V_{ud}& V_{us} & V_{ub} \\ V_{cd} & V_{cs} & V_{cb}\\ V_{td} & V_{ts} & V_{tb}\end{pmatrix}\label{MCKMM}
\end{equation}
Expanding in $\alpha$ up to $\alpha^2$ terms included, one explicitly obtains
\beq
V_{ud}=1{-}{3\alpha^{2}\over 32}{(r_{23}^{u} r_{13}^{d} {-} r_{23}^{d} r_{13}^{u})^2\over  (r_{23}^{d}r_{23}^{u})^{2} }{+}...\label{VUD}
\eeq
\beq
 V_{us}={\alpha \sqrt{3}}{ r_{13}^{u} r_{23}^{d}{-}r_{23}^{u} r_{13}^{d}\over 4  r_{23}^{d} r_{23}^{u}}{+}{\alpha^{2} \sqrt{3}\over 8}{ (r_{13}^{u} r_{23}^{d})^{2}{-}(r_{23}^{u} r_{13}^{d})^{2}\over (r_{23}^{d} r_{23}^{u})^{2}}{+}...\label{VUS}
\eeq
\beq
V_{ub}=-\alpha^{2}{r_{23}^{u} r_{13}^{d} {-} r_{23}^{d} r_{13}^{u}\over 3\sqrt{6}\, r_{23}^{d} }{+}... \label{VUB}
\eeq
\beq
 V_{cd}={\alpha \sqrt{3}}{ r_{23}^{u} r_{13}^{d}{-}r_{23}^{d} r_{13}^{u}\over 4  r_{23}^{d} r_{23}^{u}}{+}{\alpha^{2} \sqrt{3}\over 8}{ (r_{23}^{u} r_{13}^{d})^{2}{-}(r_{23}^{d} r_{13}^{u})^{2}\over (r_{23}^{d} r_{23}^{u})^{2}}{+}...\label{VCD}
\eeq
\beq
V_{cs}=1{-}{3\alpha^{2}\over 32}{(r_{23}^{u} r_{13}^{d} {-} r_{23}^{d} r_{13}^{u})^2\over (r_{23}^{d} r_{23}^{u})^{2}}-{4\alpha^{2}\over 81}(r_{23}^{d} {-} r_{23}^{u})^2{+}...\label{VCS}
\eeq
\beq
V_{cb}={36\alpha}{r_{23}^{d}{-}r_{23}^{u}\over 81\sqrt{2}}{-}{8\alpha^2}{(r_{23}^{d})^{2}{-}(r_{23}^{u})^{2}\over 81\sqrt{2} }{+} {\alpha^2\over 9} (r_{13}^{d}{-}r_{13}^{u}) {+}...\label{VCB}
\eeq
\beq
V_{td}=\alpha^{2}{r_{23}^{u} r_{13}^{d} {-} r_{23}^{u} r_{13}^{u}\over 3\sqrt{6} \,r_{23}^{u}}{+}...\label{VTD}
\eeq
\beq
V_{ts}
={36\alpha}{r_{23}^{u}{-}r_{23}^{d}\over 81\sqrt{2}}{-}{8\alpha^2}{(r_{23}^{u})^{2}{-}(r_{23}^{d})^{2}\over 81\sqrt{2} }{+} {\alpha^2\over 9} (r_{13}^{u}{-}r_{13}^{d}) {+}... \label{VTS}
\eeq
\beq
V_{tb}=1{-}4\alpha^{2}{(r_{23}^{u}{-}r_{23}^{d})^{2}\over 81}{+}...\label{VTB}
\eeq}
The elements of the matrix~(\ref{MCKMM}) show an interesting hierarchical behaviour that matches the general structure of the experimental expression of the CMK matrix, $V_{CKM}^{exp}$~\cite{CKMM} (see the numbers reported in square parentheses in eq.~(\ref{CKMFIT})).

We list here some of the features of our theoretical CKM matrix that are amazingly in agreement with the numerical pattern of $V_{CKM}^{exp}$. 

1) First of all, the overall hierarchical arrangement of the matrix elements of $V_{CKM}^{exp}$ is nicely captured by the $\alpha$ dependence of the matrix elements from~(\ref{VUD}) to~(\ref{VTB}). In fact, one can check that the further away we move from the diagonal the smaller is the size (of the modulus) of the corresponding matrix elements. This is in line with the pattern provided by the Wolfenstein parametrization~\cite{WOLF} with the identification $\lambda_{Wolf}= {\mbox{O}}(\alpha)$.

2) Up to O($\alpha^2$) included one gets the two relations
\beq
V_{ts}=-V_{cb} +{\mbox{O}}(\alpha^{3})\,,\quad V_{us}=-V_{cd} +{\mbox{O}}(\alpha^{3}) \label{VUSVCD}
\eeq
which are rather well satisfied by the  $V_{CKM}^{exp}$ entries. 

3) With an eye to the Cabibbo angle definition~\cite{Cabibbo:1963yz}, from 
\beq
V_{cd}={\alpha \sqrt{3}\over 4  r_{23}^{d} r_{23}^{u}}( r_{23}^{u} r_{13}^{d}{-}r_{23}^{d} r_{13}^{u})+{\mbox{O}}(\alpha^2) 
\label{STHETAC}
\eeq
 we get
 \begin{eqnarray}
 \sqrt{1{-}V_{cd}^2}&&\!\!=1-{1\over 2}{3\alpha^{2}\over 16 (r_{23}^{d}r_{23}^{u})^{2}}(r_{23}^{u} r_{13}^{d} - r_{23}^{d} r_{13}^{u})^2+{\mbox{O}}(\alpha^{3})\nn\\ &&\!\!=V_{ud}+{\mbox{O}}(\alpha^{3}) 
 \label{CTHETAC}
 \end{eqnarray}
 in line with the identifications $V_{cd}=\sin\theta_c +{\mbox{O}}(\alpha^2)$ and $\sqrt{1-V_{cd}^2}=V_{ud}+{\mbox{O}}(\alpha^{3}) =\cos\theta_c+{\mbox{O}}(\alpha^{3})$. This is of no surprise as our CKM matrix is  unitary. 
 
4) Relations like in 2) and 3) hold also among $V_{ts}$, $V_{tb}$ and $V_{cb}$ to the first non-trivial order in $\alpha$.
 
5) By inspection we find the ``sum rule''
\begin{equation}
\begin{split}
V_{tb}&=1-{4\over 81}\alpha^{2}(r_{23}^{u}{-}r_{23}^{d})^{2} +{\mbox{O}}(\alpha^{4})\\
&=1-(V_{ud}{-}V_{cs})+{\mbox{O}}(\alpha^{4})
\end{split} \label{SURU}
\end{equation}
which is surprisingly well fulfilled by data. 


\section{A bit of phenomenology}
\label{sec;ABOP}

We now want to analyze the numerical implications of the expressions we obtained in the previous section for quark masses (eqs.~(\ref{EVAU})-(\ref{EVAT})) and CKM matrix elements (eqs.~(\ref{VUD})-(\ref{VTB})).

This can be done in two ways. The first is to fit the four parameters $\rho^u_{13},\rho^u_{23},\rho^d_{13},\rho^d_{23}$ against the experimental values of the four quark mass ratios $m_c/m_u, m_t/m_c, m_s/m_d, m_b/m_s$ and the entries of the CKM matrix without expanding in $\alpha$.

The second consists in solving the equations for the four quark mass ratios $\lambda_{2}^u/\lambda_{1}^u, \lambda_{3}^u /\lambda_{2}^u, \lambda_{2}^d/\lambda_{1}^d, \lambda_{3}^d/ \lambda_{2}^d$ in favor of the four parameters $\rho^u_{13}, \rho^u_{23}, \rho^d_{13}, \rho^d_{23}$ and then plugging the latter back into the expressions~(\ref{VUD})-(\ref{VTB}) of the CKM entries. 
\subsection*{Fitting quark mass ratios and CKM matrix elements}
\label{sec:QMFIT}
In order to determine the best fit values of the four $\rho$ parameters, 
we choose to minimize the target function. 
\begin{widetext}
\small
\begin{equation}
F_{fit}:=\left(1{-} {|V_{us}|\over |V_{us}^{exp}|} \right)^{2} \!+\! \left(1{-} {|V_{ts}|\over |V_{ts}^{exp}|} \right)^{2}\!+\!\left(1{-} {|V_{td}|\over |V_{td}^{exp}|} \right)^{2} \!+\!\left(1{-} {\lambda_{2}^u \over \lambda_{1}^u}{m_{u}\over m_{c}} \right)^{2}+\left(1{-} {\lambda_{3}^u\over \lambda_{2}^u}{m_{c}\over m_{t}} \right)^{2}\!+\!
\left(1{-} {\lambda_{2}^d\over \lambda_{1}^d}{m_{d}\over m_{s}} \right)^{2}\!+\!\left(1{-} {\lambda_{3}^d \over \lambda_{2}^d}{m_{s}\over m_{b}} \right)^{2} \label{FPPT}
\end{equation}
\end{widetext}
where (masses are expressed in MeV and are all taken at the same renormalization point)
\beqn
&&|V_{us}^{exp}|\simeq 0.2264\,,\quad |V_{ts}^{exp}|\simeq 0.0398 \,,\quad |V_{td}^{exp}|\simeq 0.0085\nn\\
&&m_{u}=3\,,\quad m_{c}=1.27\times 10^{3}\,,\quad m_{t}=170 \times 10^{3} \nn\\
&&m_{d}=4.8\,,\quad m_{s}=95 \,,\quad m_{b}=4.3 \times 10^{3}\label{MASSES}
\eeqn
We find convenient to carry out the minimization procedure 
with respect to $\rho_{23}^{u,d}$ and $\rho_{12}^{u,d}\equiv\rho_{13}^{u,d}/\rho_{23}^{u,d}$. 
At the minimum we have the following results:

1) for the target function $F_{fit}=0.78$,

2) for the values of the fitting parameters
\small{\beq
\rho_{12}^{u}=0.003,\, \rho_{23}^{u} = 0.034\, ,\quad \rho_{12}^{d}=0.443,\, \rho_{23}^{d}= 0.150 \,, \label{RHOFIT}
\eeq}
with fit errors 
{\small\beqn
\hspace{-.6cm}&&\Delta\rho_{12}^{u}\!=\!0.002\,, \, \Delta\rho_{23}^{u}\!=\!0.018\,,\, \Delta\rho_{12}^{d}\!=\!0.22\,,\, \Delta\rho_{23}^{d}\!=\!0.09\,,
\eeqn}

3) for the entries of the CKM 
matrix
\begin{widetext}
\beq
V_{CKM}^{fit[exp]}=\left(
\begin{array}{lll}
 0.9738\,\,[0.97401\pm 0.00011]_{ud} & -0.2361\,\,[0.22636 \pm 0.00048]_{us} & -0.0020\,\,[0.00361\pm 0.0001 ]_{ub}\\
 0.2360\,\,[0.22650 \pm 0.00048]_{cd} & \,\,\,\,\,0.9730\,\,[0.97320 \pm 0.00011]_{cs} & -0.0408\,\,[0.04053\pm 0.0007]_{cb}\\
 0.0090\,\,[0.00854\pm 0.0002]_{td} & \,\,\,\,\,0.0397\,\,[0.03978\pm 0.0007]_{ts} & \,\,\,\,\,0.9992\,\,[0.999172 \pm 0.000024]_{tb}  \\
\end{array}
\right)\,,\label{CKMFIT}
\eeq
\end{widetext}
where to allow for an easy comparison with the data, we give in square parentheses the experimental values of the CKM entries with their errors~\cite{CKMM},

4) for the quark mass ratios 
{\small\beqn
\hspace{-.8cm}&&\dfrac{m_t}{m_c}\Big{|}_{fit}\!\!=\!\dfrac{\lambda_{3}^u}{\lambda_{2}^u}\Big{|}_{fit}\!\!\sim \!136\,[133], \,\,\,\,\,\,\dfrac{m_b}{m_s}\Big{|}_{fit}\!= \!\dfrac {\lambda_{3}^d}{ \lambda_{2}^d}\Big{|}_{fit}\!\!\sim 30 \, [45] \label{LAMFIT1}\\
\hspace{-.8cm}&&\dfrac{m_c}{m_u}\Big{|}_{fit}\!\!=\!\dfrac{\lambda_{2}^u}{\lambda_{1}^u}\Big{|}_{fit}\!\!\sim\! 438\,[423],\,\,\,\,\,\,\dfrac{m_s}{m_d}\Big{|}_{fit}\!=\!\dfrac{\lambda_{2}^d}{ \lambda_{1}^d}\Big{|}_{fit}\!\!\sim \! 4\,[20]
\label{LAMFIT2}
\eeqn}
with the experimental values of the quark mass ratios reported in square parentheses next to the fitted numbers.

The propagated fit errors on $V_{CKM}^{fit[exp]}$ and quark mass ratios are quite large, of the order of 50\%. 

\subsubsection{Observations}

The whole procedure and the fit results require a few comments.

1) As one can see from eqs.~(\ref{EVAU})-(\ref{EVAT}), the coefficients $\rho$ and hence the entries of the whole $V_{CKM}^{fit}$ matrix can be expressed in terms of the quark mass ratios. This important observation shows that the elements of the CKM matrix are RGI quantities because in mass ratios the QCD running factors cancel. 

2) Since everything is a function of the $\rho$ parameters (see  eq.~(\ref{MTRIAL2NUD})), in view of the scaling with the gauge coupling expressed by the eqs.~(\ref{DEFR}), the quantities expected to be O(1) are the coefficients $r$~\cite{Rossi:2024fku}. To get an estimate of the fitted $r$ values, we rely on the fact that they are RGI quantities. We can thus compute them at any scale. Since, as shown in ref.~\cite{Frezzotti:2016bes}, in the kind of bSMm's we are interested in here gauge couplings show unification, we decide to take $\alpha=\alpha_{GUT}$. At the unification scale one finds $\alpha_{GUT}\sim 1/28$ (see fig.~6 of~\cite{Frezzotti:2016bes}). According to the scaling laws~(\ref{DEFR}), from the fitted numbers~(\ref{RHOFIT}) we obtain the somewhat more ``natural'' set of O(1) values $r_{12}^{u}\sim 0.003 \times 28=0.084\, , r_{23}^{u} \sim 0.034\times 28 = 0.95$ and $r_{12}^{d}\sim 0.443\times 28= 12.4\, , r_{23}^{d}\sim 0.150\times 28=4.2$.

3) As for the quality of the fit, we observe that the elements of the CKM matrix $V_{CKM}^{fit}$ are all in good agreement with the numbers in $V_{CKM}^{exp}$ (see eq.~(\ref{CKMFIT})). Looking at eqs.~(\ref{LAMFIT1})-(\ref{LAMFIT2}) we see that this is also true for ${\lambda_{2}^u}/{\lambda_{1}^u}\sim m_c/m_u, {\lambda_{3}^u}/{\lambda_{2}^u}\sim m_t/m_c$ and ${\lambda_{3}^d}/{\lambda_{2}^d}\sim m_b/m_s$. 
Unfortunately, an inconsistent prediction is instead obtained for ${\lambda_{2}^d}/{\lambda_{1}^d}\sim m_s/m_d$. One finds ${\lambda_{2}^d}/{\lambda_{1}^d}\sim 4$ rather than the physical value $\sim 20$. In general, it looks that the down ratios are less well reproduced by the fit than the up ratios.  
\vspace{-.4cm}

\subsection*{Solving for quark mass ratios}

To get some feeling about the previous findings and in particular about the difference between the fitted number of ${\lambda_{2}^d}/{\lambda_{1}^d}$ and the physical value of $m_s/m_d$ we can offer the following argument.

With the identifications entailed by the eqs.~(\ref{IDMUCT}) and~(\ref{IDMDSB}) we can solve the four equations for the mass ratios that can be obtained from eqs.~(\ref{EVAU})-(\ref{EVAT}) in favour of four independent $\rho$ parameters and then plug these values back into our expressions~(\ref{VUD})--(\ref{VTB}) of the CKM matrix elements. 
For an easier understanding of the situation, it is convenient to expand our formulae taking into account the scaling~(\ref{DEFR}), or equivalently the fact that the $\rho$ parameters~(\ref{RHOFIT}) are all $\ll 1$. 

Identifying at leading order the quantities $\lambda^u_i,\lambda^d_i, i=1,2,3$ with the up and down quark masses, we get the relations
{\small\beqn
&&\dfrac{\rho_{13}^{u}}{\rho_{23}^{u}} =6\dfrac{m_{u}}{m_{t}} \left[\dfrac{9}{2}\dfrac{m_{c}}{m_{t}}\right]^{-1}+\ldots=\dfrac{4}{3}\dfrac{m_{u}}{m_{c}}+\ldots\label{ROU21}\\
&&\dfrac{\rho_{13}^{d}}{\rho_{23}^{d}} =6\dfrac{m_{d}}{m_{b}} \left[\dfrac{9}{2}\dfrac{m_{s}}{m_{b}}\right]^{-1}+\ldots=\dfrac{4}{3}\dfrac{m_{d}}{m_{s}}+\ldots\label{ROU31}
\eeqn}
With the help of eqs.~(\ref{ID1})--(\ref{ID6}) and the relations~(\ref{ROU21})-(\ref{ROU31}) we can express the CKM entries~(\ref{VUD})--(\ref{VTB}) in terms of quark mass ratios, obtaining
{\small
\beqn
\hspace{-.7cm}&&V_{ud}= 1{-}\dfrac{1}{6}\dfrac{m_{d}^2}{m_{s}^2}{+}...\,,\quad V_{us}=-\dfrac{1}{\sqrt{3}}\left(  \dfrac{m_{d}}{m_{s}} {-} \dfrac{m_{u}}{m_{c}} \right){+}...\label{VUD1}\\
\hspace{-.7cm}&&\qquad \qquad \quad V_{ub}= \dfrac{\sqrt{6}}{3}\left(\dfrac{m_{u}}{m_{t}}{-}\dfrac{m_{c}m_d}{m_{t}m_s}\right){+}...\label{VUB1}\\
\hspace{-.7cm}&&V_{cd}= \dfrac{1}{\sqrt{3}}\left( \dfrac{m_{d}}{m_{s}} {-} \dfrac{m_{u}}{m_{c}} \right){+}...\,,\quad V_{cb}= \sqrt{2}\,\dfrac{m_{s}}{m_{b}}{+}...\label{VUB1}\\
\hspace{-.7cm}&&\qquad \qquad \quad V_{cs}= 1{-}\dfrac{1}{6}\left(\dfrac{m_{d}}{m_{s}}\right)^2 {-}\left(\dfrac{m_{s}}{m_{b}}\right)^2{+}...\label{VUB2}\\
\hspace{-.7cm}&&V_{td}\!= \dfrac{\sqrt{6}}{3}\left(\dfrac{m_{d}}{m_{b}}{-}\dfrac{m_{s}m_u}{m_{b}m_c}\right){+}...\,,\quad V_{ts} \!=\! - \sqrt{2}\,\dfrac{m_{s}}{m_{b}}{+}...\label{VTS1}\\
\hspace{-.7cm}&& \qquad \qquad \qquad\qquad V_{tb}=1{-} \dfrac{m_{s}^2}{m^2_{b}}{+}...\label{VTB1}
\eeqn}
where dots are higher order terms in the counting~(\ref{IDMUCT})-(\ref{IDMDSB}).

Eqs.~(\ref{VUD1})--(\ref{VTB1}) are the expansions of the exact expression of the CKM entries~(\ref{VUD})-(\ref{VTB}), showing that that sinus of the Cabibbo angle is $\propto m_d/m_s$. Naturally, inserting the fitted numbers~(\ref{LAMFIT1})-(\ref{LAMFIT2})returns the matrix~(\ref{CKMFIT}). 

Unfortunately, there is a clash between the fitted and the physical value of the ratio $m_d/m_s$ which is very small and would yield 
\beq
V_{cd} \sim -V_{us} \sim\dfrac{1}{\sqrt{3}} \dfrac{m_{d}}{m_{s}}\sim 0.03\, ,\label{SC}
\eeq
a number for the sinus of the Cabibbo angle smaller by a factor $6 \div 7$ compared to experiments. 

Despite this disturbing numerical issue, we wish to stress a further nice feature of our CKM matrix construction. As we also see from the eqs.~(\ref{VUD1})--(\ref{VTB1}), in the present scheme the entries of the CKM matrix can be expressed in terms of quark mass ratios without any need to know the coefficients $C_{q^{u,d}}^{(i)}$, $i\!=\!1,2,3$, which depend on the detailed form of the $d\geq 6$ Wilson-like operators in the fundamental Lagrangian.

\vspace{.4cm}
\section{Conclusions}
\label{sec:CONC}

Dwelling on the results of refs.~\cite{Frezzotti:2014wja,Rossi:2024fku}, we have proposed a theoretical scenario that allows a physically motivated understanding of the structure of the CKM matrix~\cite{CKMM}. The approach also provides some clue about the origin of flavour.

We find that, although the fitted value of $m_d/m_s$ is in disagreement with data (it comes out too large), the numerical estimate of the entries of the CKM matrix and the three other quark mass ratios match rather well their experimental values. 

A possible way to reconcile the fitted value of $m_d/m_s$ with phenomenology could be to invoke EW corrections. Although radiative corrections to the form of the mass matrices~(\ref{MTRIAL2NUD}) do not affect the leading expressions of the quark masses~\cite{Rossi:2024fku}, they may induce higher order deformations of the ``rigid'' structure of $M^{u,d}$, possibly mitigating the problem.

In conclusion, the overall structure of the CKM matrix and the relations from~(\ref{VUSVCD}) to~(\ref{SURU}) we have derived, are robust post-dictions weakly depending on the value of $\alpha$ and the detailed expression of the $d \geq 6$ Wilson-like operators that trigger the mechanism of the NP generation of quark masses~\cite{Frezzotti:2014wja,Rossi:2024fku}.

\vspace{4 mm}
\section*{Acknowledgements} 
We thank R.\ Frezzotti for his interest in the first stages of this investigation. We are grateful to R.\ Barbieri for useful correspondence.\ GCR acknowledges partial financial support from INFN IS Lqcd123. The work of MF was supported by the European MSCA grant HORIZON-MSCA-2022-PF-01-01 ``BlackHoleChaos'' No.101105116 and partially supported by the H.F.R.I call ``Basic research Financing (Horizontal support of all Sciences)'' under the National Recovery and Resilience Plan ``Greece 2.0'' funded by the European Union ``Next Generation EU'' (H.F.R.I.) Project Number: 15384.

\bibliographystyle{apsrev4-2}

\end{document}